\documentstyle[preprint,aps]{revtex}

\tightenlines

\begin{document}
\title{Exact Renormalization of Massless QED$_{2}$}
\author{Rodolfo Casana S.$^{\text{1}}$\thanks{%
casana@ift.unesp.br} \quad and\quad\ Sebasti\~{a}o Alves Dias$^{\text{2,3}}$%
\thanks{%
tiao@cbpf.br}}
\address{$^{1}${\small Instituto de F\'{\i}sica Te\'{o}rica-UNESP}\\
{\small \ Rua Pamplona, 145, 01405-900, S\~{a}o Paulo, SP, Brazil}\\
$^{2}${\small Centro Brasileiro de Pesquisas F\'{\i}sicas, Departamento de}\\
Campos e Part\'{\i}culas\\
{\small \ Rua Xavier Sigaud, 150, 22290-180, Rio de Janeiro, RJ, Brazil}\\
{\small \ }$^{3}${\small Pontif\'{\i}cia Universidade Cat\'{o}lica do Rio de}%
\\
Janeiro, Departamento de F\'{\i}sica\\
{\small \ Rua Marqu\^{e}s de S\~{a}o Vicente, 225, 22543-900, Rio de}\\
Janeiro, RJ, Brazil {\small \ }}
\date{\today}
\maketitle
\pacs{11.15.-q, 11.15.Tk}

\begin{abstract}
We perform the {\it exact} renormalization of two-dimensional massless gauge
theories. Using these exact results we discuss the cluster property and
confinement in both the anomalous and chiral Schwinger models.
\end{abstract}

\section{Introduction}

\noindent

The gauge principle has been the most important tool to build the models
that describe successfully most of the fundamental interactions. The
possibility of a perturbative renormalization of gauge theories made this
description practical. Quantum gauge invariance was fundamental in producing
Ward or Slavnov-Taylor identities that related renormalization constants and
allowed cancellation of otherwise intractable divergences. Gauge anomalies
apparently destroy quantum gauge invariance and invalidate the perturbative
renormalization program. As almost everything that is accessible from these
theories is obtained on perturbative grounds, at least in four dimensions,
this turned cancellation of gauge anomalies almost into a new physical
principle.

In two dimensions, however, it is possible to examine in \ much more detail
the renormalization of anomalous gauge theories. Jackiw and Rajaraman showed
the quantum consistency of an anomalous gauge theory (two dimensional
massless QED with chiral fermions, also called {\it chiral Schwinger model}
(CSM)) \cite{jackiw-raja}. Other studies (\cite{faddeev}, \cite{schapo}, 
\cite{harada}) showed that the gauge anomaly, at least in this context, far
from being a sign of inconsistency, was a source of dynamic richness. The
model was shown to exhibit the dynamical mass generation phenomenon for
gauge bosons, without fermion screening or confinement, a highly desirable
characteristic in a realistic theory of weak interactions. More recently,
the conventional Schwinger model (two dimensional massless QED with Dirac
fermions) regularized in a non-gauge invariant way \cite{jackiw-johnson} (we
call it {\it anomalous Schwinger model} (ASM)) was considered as a possibly
non-confining model with dynamical mass generation as well \cite{mitra}. In
both cases, the greatest emphasis was put in the formal consistency of the
theory, in the structure of the Hilbert space, constraints, etc. No
questions about renormalization have been raised in these papers, as it was
not necessary to consider it in detail for most of these discussions. They
concentrated in aspects related to the gauge boson propagator, and bosonic
correlation functions are finite (up to regularization of the photon
self-energy to one loop).

This does not mean that the renormalization problem was absent. The need for
a fermion wave function renormalization has been recognized long time ago
both in the CSM \cite{girotti}\cite{boyanovsky} and in the ASM \cite{mitra}.
The general structure of this renormalization was studied and clarified in 
\cite{chineses}.\ It is now clear how it can be performed, in a
semi-perturbative regime \cite{ijmp}, and what is the precise structure and
origin of the divergence \cite{jpg}. But would it be possible that the {\it %
exact} renormalization of these models could be performed?

There are crucial aspects of the models (like the precise infrared behavior
of the renormalized correlation functions) that may, in principle, depend on
the details of the exact{\it \ }(here taken as the opposite of {\it %
perturbative}) renormalization. These aspects are fundamental to establish
physical properties like confinement, for example. Also, one could ask: why
should these models be called {\it exactly solvable}? In principle, because
one could compute {\it any} correlation function, bosonic or fermionic, with
an arbitrary number of external legs, {\it exactly, }or in other words, to
all orders of the coupling constant. For the bosonic ones this is true, but
this does not happen for the fermionic ones. After all, thanks to the need
of renormalization, we could only compute correlation functions {\it %
semi-perturbatively} \cite{ijmp}, and this is just a little more than what
we can obtain in four dimensional, not exactly solvable theories. Not
answering these questions is really not knowing the fermionic sector of the
theory (or knowing it very poorly, within the context of a rough
approximation).

In this paper we perform the {\it exact} renormalization of massless QED$%
_{2} $. This means the {\it exact} computation of the renormalization
constant and the definition of {\it exactly} renormalized fermion
amplitudes. We use the two-point fermion amplitude in ASM and CSM to
investigate the possible existence of asymptotic fermions and the cluster
property, thus addressing the question of confinement in these theories.
Because the calculations are quite similar for both the CSM and ASM, we will
detail our calculations in the ASM, in section 2, and only state the results
for the CSM, in section 3. We do this because, as we will show, the ASM has
much less trivial physical properties than the CSM, in the sense that two
regimes appear after renormalization, one for non-confined and one for
eventually confined fermions. In section 4 we present our conclusions.

\section{Exact Renormalization of the Anomalous Schwinger Model}

The anomalous Schwinger model is defined by the following regularized
Lagrangian density \cite{ijmp}, 
\begin{equation}
{\cal L}_{v}^{\Lambda }\left[ \psi ,\bar{\psi},A\right] =-\frac{1}{4}F_{\mu
\nu }F^{\mu \nu }-\frac{e^{2}\left( a_{v}-1\right) }{4\pi \Lambda ^{2}}%
\left( \partial \cdot A\right) ^{2}+\bar{\psi}\left( i\partial
\!\!\!/+eA\!\!\!/\right) \psi \text{,}
\end{equation}
where the parameter $\Lambda $ has the dimension of a mass, that goes to
infinity at the end of the computations, $\psi $ is a Dirac fermion in two
dimensions and $a_{v}$ is the Jackiw-Rajaraman parameter, representing
ambiguities in the regularization of the two point photon function, which is
exactly calculated, up to this regularization \cite{jackiwtopo}\cite
{tiao-linhares2}. The fermion propagator can be exactly computed and is
given by 
\begin{equation}
G_{v}^{\Lambda }\left( x-y\right) =i\exp \left( ie^{2}\int \frac{d^{2}k}{%
\left( 2\pi \right) ^{2}}f_{v}^{\Lambda }\left( k\right) \left[ 1-e^{ik\cdot
\left( x-y\right) }\right] \right) G_{F}\left( x-y\right) ,
\label{greenfunction1}
\end{equation}
where $G_{F}\left( x-y\right) $ is the free fermion propagator and 
\begin{equation}
f_{v}^{\Lambda }\left( k\right) =\frac{2\pi \Lambda ^{2}}{e^{2}\left(
a_{v}-1\right) }\frac{1}{k^{2}\left( k^{2}-\Lambda ^{2}\right) }-\frac{1}{%
k^{2}\left( k^{2}-m_{v}^{2}\right) }\text{,}
\end{equation}
$m_{v}$ being the dynamically generated mass for the photon, 
\begin{equation}
m_{v}=\frac{e^{2}}{2\pi }\left( a_{v}+1\right) \text{.}
\end{equation}
One can easily see that, when $\Lambda $ goes to infinity, the function $%
f_{v}^{\Lambda }\left( k\right) $ (which, for $k\rightarrow \infty $ behaves
as $k^{-4}$) goes to 
\begin{equation}
f_{v}\left( k\right) =-\frac{2\pi }{e^{2}\left( a_{v}-1\right) }\frac{1}{%
k^{2}}-\frac{1}{k^{2}\left( k^{2}-m_{v}^{2}\right) }\text{,}
\end{equation}
part of which behaves as $k^{-2}$, when $k\rightarrow \infty $. This induces
a logaritmic divergence in expression (\ref{greenfunction1}). The same
divergence appears in arbitrary ($n$ point) fermionic correlation functions 
\cite{chineses}. Treating this divergence in the two point fermionic
function is then enough to define completely the theory.

We begin by noticing that the two point function satisfies a Schwinger-Dyson
equation 
\begin{equation}
\left( i\partial \!\!\!/+ie^{2}\int \frac{d^{2}k}{\left( 2\pi \right) ^{2}}%
f_{v}^{\Lambda }\left( k\right) k\!\!\!/e^{ik\cdot \left( x-y\right)
}\right) G_{v}^{\Lambda }\left( x-y\right) =i\delta \left( x-y\right) \text{.%
}
\end{equation}
This can be written, in momentum space, as 
\begin{equation}
\tilde{G}_{v}^{\Lambda }\left( p\right) =\frac{i}{p\!\!\!/}-ie^{2}\int \frac{%
d^{2}k}{\left( 2\pi \right) ^{2}}f_{v}^{\Lambda }\left( k\right) \frac{1}{%
p\!\!\!/}k\!\!\!/\tilde{G}_{v}^{\Lambda }\left( p-k\right) \text{.}
\label{Schwinger-Dyson}
\end{equation}
Iterating this equation, we obtain an expression for $\tilde{G}$ as a series
in $f_{v}^{\Lambda }$ that is the semiperturbative expansion obtained in 
\cite{ijmp}, 
\begin{equation}
\tilde{G}_{v}^{\Lambda }\left( p\right) =\frac{i}{p\!\!\!/}%
+\sum\limits_{n=1}^{\infty }\tilde{G}_{v}^{\Lambda \left( n\right) }\left(
p\right) \text{,}  \label{serie1}
\end{equation}
where we are considering 
\begin{equation}
\tilde{G}_{v}^{\Lambda \left( 0\right) }\left( p\right) =\frac{i}{p\!\!\!/}%
\text{,}
\end{equation}
and $\tilde{G}_{v}^{\Lambda \left( n\right) }\left( p\right) $ is the $n$%
-loop contribution to the fermion propagator, obtained with the use of the 
{\it exact} photon propagator. We can use equation (\ref{serie1}) together
with (\ref{Schwinger-Dyson}) to obtain a recurrence relation for the
different contributions to $\tilde{G}_{v}^{\Lambda }\left( p\right) $, 
\begin{equation}
\tilde{G}_{v}^{\Lambda \left( n+1\right) }\left( p\right) =-ie^{2}\int \frac{%
d^{2}k}{\left( 2\pi \right) ^{2}}f_{v}^{\Lambda }\left( k\right) \frac{1}{%
p\!\!\!/}k\!\!\!/\tilde{G}_{v}^{\Lambda \left( n\right) }\left( p-k\right) 
\text{.}
\end{equation}
To first order, we have the following identity 
\begin{eqnarray}
\tilde{G}_{v}^{\Lambda \left( 1\right) }\left( p\right) &=&-ie^{2}\int \frac{%
d^{2}k}{\left( 2\pi \right) ^{2}}f_{v}^{\Lambda }\left( k\right) \frac{1}{%
p\!\!\!/}k\!\!\!/\tilde{G}_{v}^{\Lambda \left( 0\right) }\left( p-k\right) 
\nonumber \\
&=&-i^{2}e^{2}\int \frac{d^{2}k}{\left( 2\pi \right) ^{2}}f_{v}^{\Lambda
}\left( k\right) \frac{1}{p\!\!\!/}k\!\!\!/\frac{1}{p\!\!\!/-k\!\!\!/} 
\nonumber \\
&=&-ie^{2}\int \frac{d^{2}k}{\left( 2\pi \right) ^{2}}f_{v}^{\Lambda }\left(
k\right) \left( \frac{i}{p\!\!\!/}-\frac{i}{p\!\!\!/-k\!\!\!/}\right) 
\nonumber \\
&=&ie^{2}\int \frac{d^{2}k}{\left( 2\pi \right) ^{2}}f_{v}^{\Lambda }\left(
k\right) \left( \tilde{G}_{v}^{\Lambda \left( 0\right) }\left( p\right) -%
\tilde{G}_{v}^{\Lambda \left( 0\right) }\left( p-k\right) \right) \text{,}
\end{eqnarray}
that we can easily generalize, by induction, to all orders 
\begin{equation}
\tilde{G}_{v}^{\Lambda \left( n+1\right) }\left( p\right) =\frac{ie^{2}}{n+1}%
\int \frac{d^{2}k}{\left( 2\pi \right) ^{2}}f_{v}^{\Lambda }\left( k\right)
\left( \tilde{G}_{v}^{\Lambda \left( n\right) }\left( p\right) -\tilde{G}%
_{v}^{\Lambda \left( n\right) }\left( p-k\right) \right) \text{.}
\label{recorrencia1}
\end{equation}
This expression is the basis of the solution for the Fourier transform of
the propagator. It is analogous to the one obtained in \cite{radozycki} in
the case $a_{v}=1$ (in which the ASM reduces to the conventional Schwinger
model, thanks to preservation of intermediate gauge invariance). It has its
origin in the fact that breaking gauge symmetry in intermediate steps (we
mean, by regularization) does not affect the purely fermionic Ward
identities.

The integral of the first term in (\ref{recorrencia1}) can be performed
easily. Defining it as $I_{v}^{\Lambda }$, we obtain 
\begin{equation}
I_{v}^{\Lambda }=\int \frac{d^{2}k}{\left( 2\pi \right) ^{2}}f_{v}^{\Lambda
}\left( k\right) =\frac{i}{2e^{2}\left( a_{v}-1\right) }\ln \left( \frac{%
\Lambda ^{2}}{-i\varepsilon }\right) -\frac{i}{2e^{2}\left( a_{v}+1\right) }%
\ln \left( \frac{m^{2}}{-ie}\right) \text{.}
\end{equation}
It is then easy to obtain, again by induction, the general expression for
the $n$-loop contribution for the fermion propagator in terms of the $0$%
-loop order, 
\begin{eqnarray}
\tilde{G}_{v}^{\Lambda \left( n\right) }\left( p\right) &=&\left(
ie^{2}\right) ^{n}\sum\limits_{j=0}^{n}\frac{\left( -1\right) ^{j}}{j!}\frac{%
\left( I_{v}^{\Lambda }\right) ^{n-j}}{\left( n-j\right) !}\int \frac{%
d^{2}k_{1}}{\left( 2\pi \right) ^{2}}f_{v}^{\Lambda }\left( k_{1}\right)
\int \frac{d^{2}k_{2}}{\left( 2\pi \right) ^{2}}f_{v}^{\Lambda }\left(
k_{2}\right) ...  \nonumber \\
&&...\int \frac{d^{2}k_{j}}{\left( 2\pi \right) ^{2}}f_{v}^{\Lambda }\left(
k_{j}\right) \tilde{G}^{\left( 0\right) }\left(
p-k_{1}-k_{2}-...-k_{j}\right) \text{.}
\end{eqnarray}
Inserting this formula in the expression for the complete fermionic
propagator, we obtain 
\begin{eqnarray}
\tilde{G}_{v}^{\Lambda } &=&\exp \left( ie^{2}I_{v}^{\Lambda }\right)
\sum\limits_{n=0}^{\infty }\frac{\left( -ie^{2}\right) ^{n}}{n!}\int \frac{%
d^{2}k_{1}}{\left( 2\pi \right) ^{2}}f_{v}^{\Lambda }\left( k_{1}\right)
\int \frac{d^{2}k_{2}}{\left( 2\pi \right) ^{2}}f_{v}^{\Lambda }\left(
k_{2}\right) ...  \nonumber \\
&&...\int \frac{d^{2}k_{n}}{\left( 2\pi \right) ^{2}}f_{v}^{\Lambda }\left(
k_{n}\right) \tilde{G}^{\left( 0\right) }\left(
p-k_{1}-k_{2}-...-k_{n}\right) \text{.}  \label{GemtermosdeG0-1}
\end{eqnarray}
Now we perform the $k_{n}$ integrations. To do this, we use Schwinger
parametrization for the functions $f_{v}^{\Lambda }\left( k\right) $, 
\begin{equation}
f_{v}^{\Lambda }\left( k\right) =\frac{2\pi i}{e^{2}}\int_{0}^{\infty
}d\alpha \bar{f}_{v}^{\Lambda }\left( \alpha \right) e^{i\alpha
k^{2}-\varepsilon \alpha }\text{,}
\end{equation}
with 
\begin{equation}
\bar{f}_{v}^{\Lambda }\left( \alpha \right) =\frac{1-e^{-i\alpha \Lambda
^{2}}}{a_{v}-1}-\frac{1-e^{-i\alpha m_{v}^{2}}}{a_{v}+1}\text{,}
\end{equation}
and for the free propagator 
\begin{equation}
\tilde{G}^{\left( 0\right) }\left( p-k\right) =\frac{i}{p\!\!\!/-k\!\!\!/}=%
\frac{1}{2i}\gamma ^{\mu }\frac{\partial }{\partial p^{\mu }}%
\int_{0}^{\infty }\frac{d\beta }{\beta }e^{i\beta \left( p-k\right)
^{2}-\varepsilon \beta }\text{.}
\end{equation}
Integrating over the $k_{n}$, we obtain 
\begin{eqnarray}
\tilde{G}_{v}^{\Lambda } &=&\frac{i}{p\!\!\!/}\exp \left(
ie^{2}I_{v}^{\Lambda }\right) \sum\limits_{n=0}^{\infty }\frac{1}{2^{n}n!}%
\int_{0}^{\infty }\frac{d\alpha _{1}}{\alpha _{1}}\bar{f}_{v}^{\Lambda
}\left( \alpha _{1}\right) \int_{0}^{\infty }\frac{d\alpha _{2}}{\alpha _{2}}%
\bar{f}_{v}^{\Lambda }\left( \alpha _{2}\right) ...  \nonumber \\
&&...\int_{0}^{\infty }\frac{d\alpha _{n}}{\alpha _{n}}\bar{f}_{v}^{\Lambda
}\left( \alpha _{n}\right) \int_{0}^{\infty }\frac{d\beta }{\beta ^{2}}%
\left( \frac{1}{\beta }+\sum\limits_{j=1}^{n}\frac{1}{\alpha _{j}}\right)
^{-2}\times  \nonumber \\
&&\times \exp \left( \frac{ip^{2}}{\frac{1}{\beta }+\sum\nolimits_{j=1}^{n}%
\frac{1}{\alpha _{j}}}-\varepsilon \beta -\varepsilon
\sum\limits_{j=1}^{n}\alpha _{j}\right) \text{.}  \label{alpha1}
\end{eqnarray}
The $\beta $-integration can be done exactly, 
\begin{equation}
\int_{0}^{\infty }\frac{d\beta }{\beta ^{2}}\frac{\exp \left( \frac{ip^{2}}{%
\frac{1}{\beta }+\sum\nolimits_{j=1}^{n}\frac{1}{\alpha _{j}}}-\varepsilon
\beta \right) }{\left( \frac{1}{\beta }+\sum\limits_{j=1}^{n}\frac{1}{\alpha
_{j}}\right) ^{2}}=\frac{i}{p^{2}}\left[ 1-\exp \left( \frac{ip^{2}}{%
\sum\nolimits_{j=1}^{n}\frac{1}{\alpha _{j}}}\right) \right] \text{,}
\end{equation}
so that, inserting it back in (\ref{alpha1}) and using the identity, valid
for $z>0$, 
\begin{equation}
1-e^{-1/4z}=\int_{0}^{\infty }d\eta J_{1}\left( \eta \right) e^{-z\eta ^{2}}%
\text{,}
\end{equation}
where $J_{1}\left( \eta \right) $ is the first class Bessel function, we
obtain 
\begin{equation}
\tilde{G}_{v}^{\Lambda }\left( p\right) =\frac{i}{p\!\!\!/}\exp \left(
ie^{2}I_{v}^{\Lambda }\right) \int_{0}^{\infty }d\eta J_{1}\left( \eta
\right) \exp \left[ \frac{1}{2}\int_{0}^{\infty }\frac{d\alpha }{\alpha }%
\bar{f}_{v}^{\Lambda }\left( \alpha \right) e^{-\varepsilon \alpha -i\eta
^{2}/4p^{2}\alpha }\right] \text{.}
\end{equation}
Now, using the identity below, involving the second class modified Bessel
function $K_{0}$, 
\begin{equation}
\int_{0}^{\infty }\frac{d\alpha }{\alpha }\left( 1-e^{-i\alpha m^{2}}\right)
e^{-\varepsilon \alpha -i\eta ^{2}/4\alpha }=2K_{0}\left( \sqrt{i\varepsilon
\eta ^{2}}\right) -2K_{0}\left( \sqrt{-m^{2}\eta ^{2}}\right) \text{,}
\end{equation}
we get 
\begin{equation}
\tilde{G}_{v}^{\Lambda }\left( p^{2}\right) =\frac{i}{p\!\!\!/}\left( \frac{%
\Lambda ^{2}}{m_{v}^{2}}\right) ^{1/\left( 1-a_{v}^{2}\right) }\tilde{G}%
_{v}\left( p^{2}\right) \text{,}
\end{equation}
with 
\begin{equation}
\tilde{G}_{v}\left( p^{2}\right) =\int_{0}^{\infty }d\eta J_{1}\left( \eta
\right) \left[ -\frac{\tilde{m}_{v}^{2}\eta ^{2}}{p^{2}}\right] ^{1/\left(
1-a_{v}^{2}\right) }\exp \left\{ \frac{1}{1+a_{v}}K_{0}\left( \sqrt{-\frac{%
m_{v}^{2}\eta ^{2}}{p^{2}}}\right) \right\} \text{,}  \label{G(p)}
\end{equation}
where $\tilde{m}_{v}^{2}=\left( m_{v}^{2}e^{2\gamma _{E}}\right) /4$. This
expression is enough to compute the exact fermion wave function
renormalization needed to renormalize the theory. Proceeding as is usual, we
impose a suitable renormalization condition on the 1PI renormalized two
point function 
\begin{equation}
\left. \tilde{\Gamma}_{v}^{R}\left( p\right) \right| _{p\!\!\!/=\mu
/\!\!\!}=\mu \!\!\!/\text{,\qquad }\tilde{\Gamma}_{v}^{R}\left( p\right) 
\tilde{G}_{v}^{R}\left( p\right) =i\text{,\qquad }\tilde{G}_{v}^{R}\left(
p\right) =Z_{\psi }^{v}{}^{-1}\tilde{G}_{v}^{\Lambda }\left( p\right) \text{.%
}
\end{equation}
This shows that 
\begin{equation}
Z_{\psi }^{v}=\left( \frac{\Lambda ^{2}}{m_{v}^{2}}\right) ^{1/2\left(
1-a_{v}\right) }\tilde{G}_{v}\left( \mu ^{2}\right) \text{.}
\end{equation}
So, the expression for the exact renormalized two point function (and its
1PI counterpart) is 
\begin{equation}
\tilde{G}_{v}^{R}\left( p\right) =\frac{i}{p\!\!\!/}\frac{\tilde{G}%
_{v}\left( p^{2}\right) }{\tilde{G}_{v}\left( \mu ^{2}\right) }\text{,\qquad 
}\tilde{\Gamma}_{v}^{R}\left( p\right) =p\!\!\!/\frac{\tilde{G}_{v}\left(
\mu ^{2}\right) }{\tilde{G}_{v}\left( p^{2}\right) }\text{.}
\end{equation}
Now it is easy to obtain, taking into account the properties of $K_{0}\left(
\eta \right) $ and $J_{1}\left( \eta \right) $, when $\eta \rightarrow 0$,
the exact behavior of the fermion propagator, when $p\rightarrow 0$, 
\begin{equation}
\tilde{G}_{v}^{R}\left( p\right) \approx \frac{i}{p\!\!\!/}\left( \frac{p^{2}%
}{\tilde{m}_{v}^{2}}\right) ^{1/\left( a_{v}^{2}-1\right) }\text{.}
\end{equation}

For $a>1$, the propagator has, at most, a pole with null residue in $p^{2}=0$%
, what indicates that there are no single particle states, as it happens in
the conventional Schwinger model \cite{schwingerpolo}, in the Thirring model 
\cite{thirringpolo} and in the Schr\"{o}er model \cite{schroer}. In
particular we stress that this behavior is the same as that of the {\it exact%
} propagator of the Thirring model, if we perform a suitable correspondence
between the coupling constants. The Thirring model is characterized by the
following Lagrangian density 
\begin{equation}
{\cal L}_{Th}=\bar{\psi}i\partial \!\!\!/\psi -\frac{g}{2}\left( \bar{\psi}%
\gamma ^{\mu }\psi \right) ^{2}\text{,}
\end{equation}
with a dimensionless coupling constant $g>0$. It is possible, by using the
same techniques developed in this section, to find the exact renormalized
propagator of this model 
\begin{equation}
\tilde{G}_{Th}^{R}\left( p\right) =\frac{i}{p\!\!\!/}\left( \frac{p^{2}}{\mu
^{2}}\right) ^{\bar{g}^{2}/\left( 1+2\bar{g}\right) }\text{,\qquad }\bar{g}%
=g/2\pi \text{.}
\end{equation}
If we choose $a_{v}>1$, such that 
\begin{equation}
a_{v}=1+\frac{1}{\bar{g}}\text{,}
\end{equation}
we see that the behavior at low momentum of the ASM propagator is the same
as that of the exact Thirring model propagator.

In the interval $-1\leq a_{v}<1$, there are no simple poles at $p^{2}=0$,
what means that there are no asymptotic one particle states in this case
too. However, it exhibits poles of order greater than one at $p^{2}=0$, a
symptom of infrared slavery \cite{morchio-strocchi}, what can mean that
fermions are confined in this sector. The value of $a_{v}$, in this case,
determines the order of the pole.

For $a_{v}=1$, the infrared behavior is 
\begin{equation}
\tilde{G}_{a_{v}=1}\left( p\right) \approx \frac{i}{p\!\!\!/}\left( \frac{%
\tilde{e}^{2}/\pi }{p^{2}}\right) ^{1/4}\text{,}  \label{infra}
\end{equation}
which is a well known result \cite{radozycki}\cite{stam}. One can see
directly that the residue of the pole at $p^{2}=0$ is null, which means the
absence of asymptotic fermionic states. The question of whether this means
confinement or not is more subtle, and requires analysis of other kinds of
correlation functions, including the bosonic ones.

More information can be obtained if we come back to coordinate space, where
we will see that there is an even more explicit expression for the
renormalized fermion propagator. In doing that, first we continue the
propagator to Euclidean space making $p_{0}=ip_{2}$, $\gamma _{0}=i\gamma
_{2}$. Then, in expression (\ref{G(p)}) after renormalization, we perform
the following change of variables, remembering that $p=\sqrt{p_{\mu }p^{\mu }%
}$, 
\begin{equation}
\eta =xp\text{,}
\end{equation}
and use the identity 
\begin{equation}
\gamma _{\mu }\frac{\partial }{\partial p_{\mu }}J_{0}\left( xp\right) =-x%
\frac{p\!\!\!/}{p}J_{1}\left( xp\right) \text{,}
\end{equation}
to obtain 
\begin{eqnarray}
\tilde{G}_{v}^{R}{}_{E}\left( p\right) &=&\frac{i}{\tilde{G}_{v}\left( \mu
^{2}\right) }\gamma _{\mu }\frac{\partial }{\partial p_{\mu }}%
\int_{0}^{\infty }dx\frac{J_{0}\left( xp\right) }{x}\left[ \tilde{m}%
_{v}^{2}x^{2}\right] ^{1/\left( 1-a_{v}^{2}\right) }  \nonumber \\
&&\times \exp \left\{ \frac{1}{1+a_{v}}K_{0}\left( \sqrt{m_{v}^{2}x^{2}}%
\right) \right\} \text{.}
\end{eqnarray}
Using now the following representation for $J_{0}$%
\begin{equation}
J_{0}\left( xp\right) =\frac{1}{2\pi }\int_{0}^{2\pi }d\theta e^{-ixp\cos
\left( \theta -\alpha \right) }\text{,}
\end{equation}
with $\alpha =\arctan \left( p_{2}/p_{1}\right) $, and defining $x_{1}=x\cos
\theta $ and $x_{2}=x\sin \theta $, we see that $xp\cos \left( \theta
-\alpha \right) =x\cdot p$. So, 
\begin{eqnarray}
\tilde{G}_{v}^{R}{}_{E}\left( p\right) &=&\frac{i/2\pi }{\tilde{G}_{v}\left(
\mu ^{2}\right) }\int_{0}^{\infty }d^{2}xe^{-ix\cdot p}\frac{x\!\!\!/\!\!\!}{%
x^{2}}\left[ \tilde{m}_{v}^{2}x^{2}\right] ^{1/\left( 1-a_{v}^{2}\right) } 
\nonumber \\
&&\times \exp \left\{ \frac{1}{1+a_{v}}K_{0}\left( \sqrt{m_{v}^{2}x^{2}}%
\right) \right\} \text{.}
\end{eqnarray}
Continuing back to Minkowski space, 
\begin{eqnarray}
\tilde{G}_{v}^{R}{}\left( p\right) &=&-\frac{i/2\pi }{\tilde{G}_{v}\left(
\mu ^{2}\right) }\int_{0}^{\infty }d^{2}xe^{ix\cdot p}\frac{x\!\!\!/\!\!\!}{%
x^{2}}\left[ -\tilde{m}_{v}^{2}x^{2}\right] ^{1/\left( 1-a_{v}^{2}\right) } 
\nonumber \\
&&\times \exp \left\{ \frac{1}{1+a_{v}}K_{0}\left( \sqrt{-m_{v}^{2}x^{2}}%
\right) \right\}
\end{eqnarray}
we can now clearly read the expression for the renormalized propagator in
coordinate space, for spacelike $x^{\mu }$, 
\begin{equation}
G_{v}^{R}\left( x\right) =\frac{i}{\tilde{G}_{v}\left( \mu ^{2}\right) }\exp
\left\{ \frac{1}{1-a_{v}^{2}}\ln \left( -\tilde{m}_{v}^{2}x^{2}\right) +%
\frac{1}{1+a_{v}}K_{0}\left( \sqrt{-m_{v}^{2}x^{2}}\right) \right\}
G_{F}\left( x\right) \text{.}
\end{equation}
It is easy to obtain the expression valid for timelike $x^{\mu }$, again by
analytic continuation, remembering the continuations of $K_{0}$ and of the
logarithm, 
\begin{eqnarray}
K_{0}\left( \sqrt{-m_{v}^{2}}\right) &\rightarrow &\frac{i\pi }{2}%
H_{0}^{\left( 1\right) }\left( \sqrt{m_{v}^{2}x^{2}}\right) \text{,\qquad }%
x^{2}>0 \\
\ln \left( -\tilde{m}_{v}^{2}x^{2}\right) &\rightarrow &-i\pi +\ln \left( 
\tilde{m}_{v}^{2}x^{2}\right) \text{,\qquad\ \ \ \ }x^{2}>0
\end{eqnarray}
where $H_{0}^{\left( 1\right) }$ is the first class Hankel function. The
expression below allows us to analyze the cluster property of the propagator
explicitly: 
\begin{eqnarray}
G_{v}^{R}\left( x,y\right) &=&\frac{i}{\tilde{G}_{v}\left( \mu ^{2}\right) }%
\left[ -\tilde{m}_{v}^{2}\left( x-y\right) ^{2}\right] ^{1/\left(
1-a_{v}^{2}\right) }  \nonumber \\
&&\times \exp \left\{ \frac{1}{1+a_{v}}K_{0}\left( \sqrt{-m_{v}^{2}\left(
x-y\right) ^{2}}\right) \right\} G_{F}\left( x-y\right) \text{.}
\end{eqnarray}
Doing $x\rightarrow x+\lambda \eta $, while keeping $y$ fixed, and then $%
\lambda \rightarrow +\infty $ we obtain 
\begin{equation}
G_{v}^{R}\left( x+\lambda \eta ,y\right) 
\mathrel{\mathop{\longrightarrow }\limits_{\lambda \rightarrow \infty }}%
\frac{i}{\tilde{G}_{v}\left( \mu ^{2}\right) }\lambda ^{\left(
1+a_{v}^{2}\right) /\left( 1-a_{v}^{2}\right) }\left[ -\tilde{m}_{v}^{2}\eta
^{2}\right] ^{1/\left( 1-a_{v}^{2}\right) }G_{F}\left( \eta \right) \text{,}
\end{equation}
which means 
\begin{equation}
G_{v}^{R}\left( x+\lambda \eta ,y\right) 
\mathrel{\mathop{\longrightarrow }\limits_{\lambda \rightarrow \infty }}%
\left\{ 
\begin{array}{c}
\infty \text{,\qquad }-1\leq a_{v}<1 \\ 
0\text{,\qquad \qquad\ \ \ \ }a_{v}>1
\end{array}
\right. \text{.}
\end{equation}
So, in the interval $-1\leq a_{v}<1$ the cluster property is lost, while it
is apparently maintained for $a_{v}>1$ (the decision about this last fact
depends obviously on the analysis of the behavior of arbitrary correlation
functions). This result, and all the others in this section, would be
impossible to be guessed without the detailed analysis of the limits, that
became possible thanks to the exact expressions furnished after
renormalization. It can be easily seen also that the previous analysis in
momentum space is completely consistent with the results obtained in
coordinate space.

We can also can compute the short distance behavior of the propagator 
\begin{equation}
G_{v}^{R}\left( x\right) 
\mathrel{\mathop{\longrightarrow }\limits_{x^{2}\rightarrow 0}}%
\left[ -\tilde{m}_{v}^{2}x^{2}\right] ^{1/2\left( 1-a_{v}\right)
}G_{F}\left( x\right) \approx \left| x\right| ^{a_{v}/\left( 1-a_{v}\right) }%
\text{,}
\end{equation}
which shows explicitly the anomalous dimension under scaling (anomalous in
the sense of being distinct from the scaling dimension of the free
propagator). Only the limit $a_{v}\rightarrow +\infty $ produces a scaling
behavior similar to the free fermion propagator ($\left| x\right| ^{-1}$).

At this point it is worth commenting on a denomination usually given in the
literature \cite{mitra} for the ASM: it is called also {\it non-confining
Schwinger model}. Mitra and Rahaman recognized the need for a
renormalization, but did not performed it in detail. In the light of the
previous discussion, we see that there are reasons to believe that there is
fermion confinement (thanks to infrared slavery and violation of cluster
property, both indicated by our results) when the parameter $a_{v}$ lies in
the range $[-1,1)$. It is this behavior that makes this terminology
inadequate for all values of $\ a_{v}$. Again, before a detailed
consideration of renormalization, it is not possible to say that the model
is ``confining'' or ``non-confining''.

\section{Exact Renormalization of the Chiral Schwinger Model}

As we said in the introduction the renormalization procedure is quite
similar to that of the ASM. So, we will present the modifications that
appear thanks to the particular structure of the CSM and directly discuss
the implications of exact renormalization.

The regularized Lagrangian is given by 
\begin{equation}
{\cal L}_{v}^{\Lambda }\left[ \psi ,\bar{\psi},A\right] =-\frac{1}{4}F_{\mu
\nu }F^{\mu \nu }-\frac{e^{2}\left( a_{c}-1\right) }{8\pi \Lambda ^{2}}%
\left( \partial \cdot A\right) ^{2}+\bar{\psi}\left( i\partial
\!\!\!/+eA\!\!\!/P_{+}\right) \psi \text{,}
\end{equation}
where $P_{+}=\left( 1+\gamma _{5}\right) /2$, and we work with Dirac
fermions in order to have the eigenvalue problem for the determinant of the
Dirac operator well defined. The regularized fermion propagator is 
\begin{eqnarray}
G_{v}^{\Lambda }\left( x-y\right) &=&i\exp \left( ie^{2}\int \frac{d^{2}k}{%
\left( 2\pi \right) ^{2}}f_{c}^{\Lambda }\left( k\right) \left[ 1-e^{ik\cdot
\left( x-y\right) }\right] \right) P_{+}G_{F}\left( x-y\right)  \nonumber \\
&&+iP_{-}G_{F}\left( x-y\right) \text{,}
\end{eqnarray}
where now, 
\begin{eqnarray}
f_{c}^{\Lambda }\left( k\right) &=&-\frac{\left( \Lambda ^{2}\frac{a_{c}}{%
a_{c}-1}-\omega _{\Lambda }^{2}\right) ^{2}}{\omega _{\Lambda }^{2}\left(
\omega _{\Lambda }^{2}-\omega _{m}^{2}\right) }\frac{1}{\left( k^{2}-\Lambda
^{2}\right) \left( k^{2}-\omega _{\Lambda }^{2}\right) }  \nonumber \\
&&+\frac{\left( \Lambda ^{2}\frac{a_{c}}{a_{c}-1}-\omega _{m}^{2}\right) ^{2}%
}{\omega _{m}^{2}\left( \omega _{\Lambda }^{2}-\omega _{m}^{2}\right) }\frac{%
1}{\left( k^{2}-\Lambda ^{2}\right) \left( k^{2}-\omega _{m}^{2}\right) }
\end{eqnarray}
with $\omega _{\Lambda }$ and $\omega _{m}$ satisfying the equations 
\begin{eqnarray}
\omega _{\Lambda }^{2}+\omega _{m}^{2} &=&\Lambda ^{2}+\frac{e^{2}\left(
a_{c}+1\right) }{4\pi }\text{,}  \nonumber \\
\omega _{\Lambda }^{2}\omega _{m}^{2} &=&\Lambda ^{2}m_{c}^{2}\text{.}
\end{eqnarray}
The parameter $m_{c}$ is given by 
\begin{equation}
m_{c}^{2}=\frac{e^{2}}{4\pi }\frac{a_{c}^{2}}{a_{c}-1}\text{,}
\end{equation}
and is the usual dynamically generated mass for the photon in this model 
\cite{jackiw-raja}. The Dyson-Schwinger equation satisfied by the right
piece of the propagator $G_{c}^{\Lambda }{}^{+}\left( x\right)
=P_{+}G_{c}^{\Lambda }\left( x\right) $ is, in coordinate space, 
\begin{equation}
\left( i\partial \!\!\!/+ie^{2}\int \frac{d^{2}k}{\left( 2\pi \right) ^{2}}%
f_{c}^{\Lambda }\left( k\right) k\!\!\!/e^{ik\cdot \left( x-y\right)
}\right) G_{c}^{\Lambda }\left( x-y\right) =iP_{-}\delta \left( x-y\right) 
\text{,}
\end{equation}
and, in momentum space 
\begin{equation}
\tilde{G}_{c}^{\Lambda }{}^{+}\left( p\right) =P_{+}\frac{i}{p\!\!\!/}%
-ie^{2}\int \frac{d^{2}k}{\left( 2\pi \right) ^{2}}f_{c}^{\Lambda }\left(
k\right) \frac{1}{p\!\!\!/}k\!\!\!/\tilde{G}_{c}^{\Lambda }{}^{+}\left(
p-k\right) \text{.}
\end{equation}
Again, this equation can be solved recursively in the same way that we did
in Section 2. We arrive at an expression similar to equation (\ref
{GemtermosdeG0-1}), with an $\bar{f}_{c}^{\Lambda }\left( \alpha \right) $
given by 
\begin{eqnarray}
\bar{f}_{v}^{\Lambda }\left( \alpha \right) &=&\frac{e^{2}}{2\pi }\frac{%
\left( \Lambda ^{2}\frac{a_{c}}{a_{c}-1}-\omega _{\Lambda }^{2}\right) ^{2}}{%
\omega _{\Lambda }^{2}\left( \omega _{\Lambda }^{2}-\omega _{m}^{2}\right) }%
\frac{e^{-i\alpha \Lambda ^{2}}-e^{-i\alpha \omega _{\Lambda }^{2}}}{\Lambda
^{2}-\omega _{\Lambda }^{2}}  \nonumber \\
&&-\frac{e^{2}}{2\pi }\frac{\left( \Lambda ^{2}\frac{a_{c}}{a_{c}-1}-\omega
_{m}^{2}\right) ^{2}}{\omega _{m}^{2}\left( \omega _{\Lambda }^{2}-\omega
_{m}^{2}\right) }\frac{e^{-i\alpha \Lambda ^{2}}-e^{-i\alpha \omega _{m}^{2}}%
}{\Lambda ^{2}-\omega _{m}^{2}}.
\end{eqnarray}
Following the same steps indicated before we arrive at 
\begin{equation}
\tilde{G}_{c}^{\Lambda }{}^{+}\left( p\right) =P_{+}\frac{i}{p\!\!\!/}\left( 
\frac{\Lambda ^{2}}{m_{c}^{2}}\right) ^{1/\left( 1-a_{c}\right) }\tilde{G}%
_{c}\left( p^{2}\right) \text{,}
\end{equation}
with 
\begin{equation}
\tilde{G}_{c}\left( p^{2}\right) =\int_{0}^{\infty }d\eta J_{1}\left( \eta
\right) \exp \left\{ \frac{2}{a_{c}-1}K_{0}\left( \sqrt{-\frac{m_{c}^{2}\eta
^{2}}{p^{2}}}\right) \right\} \text{.}
\end{equation}
Now we use the relation between renormalized and bare quantities ($\tilde{G}%
_{c}^{R}{}^{+}\left( p\right) =Z_{\psi }^{c}{}^{-1}\tilde{G}_{c}^{\Lambda
}{}^{+}\left( p\right) $; $\tilde{\Gamma}_{c}^{R}{}^{+}\left( p\right)
=Z_{\psi }^{c}{}\tilde{\Gamma}_{c}^{\Lambda }{}^{+}\left( p\right) $) and
the renormalization condition, 
\begin{equation}
\left. \tilde{\Gamma}_{c}^{R}{}^{+}\left( p\right) \right| _{p\!\!\!/=\mu
\!\!\!/}=\mu \!\!\!/P_{+}\text{,}
\end{equation}
to fix $Z_{\psi }^{c}$, 
\begin{equation}
Z_{\psi }^{c}=\left( \frac{\Lambda ^{2}}{m_{c}^{2}}\right) ^{1/\left(
1-a_{c}\right) }\tilde{G}_{c}\left( \mu ^{2}\right) \text{.}
\end{equation}
This gives us immediately the renormalized two-point functions 
\begin{equation}
\tilde{G}_{c}^{R}{}^{+}\left( p\right) =P_{+}\frac{i}{p\!\!\!/}\frac{\tilde{G%
}_{c}\left( p^{2}\right) }{\tilde{G}_{c}\left( \mu ^{2}\right) }\text{%
,\qquad }\tilde{\Gamma}_{c}^{R}{}^{+}\left( p\right) =p\!\!\!/P_{+}\frac{%
\tilde{G}_{c}\left( \mu ^{2}\right) }{\tilde{G}_{c}\left( p^{2}\right) }%
\text{.}
\end{equation}
Here, also, it is possible to come back to coordinate space to obtain 
\begin{equation}
G_{c}^{R}{}^{+}\left( x\right) =\frac{i}{\tilde{G}_{c}\left( \mu ^{2}\right) 
}\exp \left\{ \frac{2}{a_{c}-1}K_{0}\left( \sqrt{-m_{c}^{2}x^{2}}\right)
\right\} P_{+}G_{F}\left( x\right) \text{,}
\end{equation}
where $x^{\mu }$ is spacelike.

This propagator has the following behavior for $x^{2}\rightarrow \infty $, 
\begin{equation}
G_{c}^{R}{}^{+}\left( x\right) 
\mathrel{\mathop{\longrightarrow }\limits_{x^{2}\rightarrow \infty }}%
P_{+}G_{F}\left( x\right) \text{,}
\end{equation}
as is well known in the conventional analysis of the CSM \cite{girotti}. The
cluster property is preserved and there exist asymptotic fermions. There are
not two different regimes, as in the case of the ASM, and the parameter $%
a_{c}$ represents only an ambiguity in the mass of the photon. Our results
confirm the ones known in literature, but with a much higher degree of rigor.

A final remark concerns the short distance behavior of the propagator, 
\begin{equation}
G_{c}^{R}{}^{+}\left( x\right) 
\mathrel{\mathop{\longrightarrow }\limits_{x^{2}\rightarrow 0}}%
\left[ -\tilde{m}_{c}^{2}x^{2}\right] ^{1/\left( 1-a_{c}\right)
}P_{+}G_{F}\left( x\right) \text{,}
\end{equation}
that again (as in the case of the ASM) reveals an anomalous scaling
dimension. This, however, does not prevents the existence of asymptotic
fermions.

\section{Conclusions and Perspectives}

Both CSM and ASM have an extremely similar divergence structure and can be
regularized by the introduction of a term that is formally similar to a
gauge fixing term with an infinite gauge parameter (in fact it has its
origin in a Pauli-Villars regularization of the propagator of the
longitudinal part of the gauge field \cite{ijmp}). The origin of the
divergence is the lack of intermediate gauge invariance, which forces one to
take into account the longitudinal degree of freedom or, in the gauge
invariant formalism, the Wess-Zumino field. Both, when {\it exactly}
integrated, give rise to a divergence in fermionic correlation functions,
thanks to their interaction with singular currents. If one uses the {\it %
complete} photon propagator in the loopwise expansion of the fermionic
correlation functions, then the divergence is under control and can be
renormalized in a quite conventional way, defining what we called a {\it %
semi-perturbative} approach.

The sensation that the {\it exact} renormalization of both ASM and CSM was
possible remained. The divergence could be explicitly computed, in its
regularized form, non-perturbatively. The main obstacle to remove it is that
the renormalization procedure in coordinate space is not well known.
Usually, one goes to momentum space, where it is clear what are the
renormalization conditions to be imposed in order to fix systematically the
form of the renormalization constants. The point is that bosonization
(necessary to establish the precise form of the divergence and ultimately to
exactly solve the model) is only known in coordinate space. One way to go to
momentum space would be in a semi-perturbative way, as we have shown \cite
{ijmp}, but then one looses the complete information about the divergence,
present in coordinate space.

In this paper we performed the {\it exact} renormalization of some gauge
theories by performing {\it exactly} the necessary Fourier transform.
Although exact renormalizations have been already achieved, within the
context and using specific features of supersymmetric theories \cite{shifman}%
, this is the first time (to our knowledge) that this is done outside this
context. We see that the result for the ASM depends strongly on the value
assumed by the Jackiw-Rajaraman parameter $a_{v}$. The precise form of this
dependence could not be guessed perturbatively and the value $a_{v}=1$ seems
to be critical to characterize confinement or not. To answer this question
completely we should compute generalized fermionic correlation functions and
verify the cluster property in connection to the appearance of $\theta $%
-vacua. A analysis of the Wilson loop, for the renormalized theory, can help
to clarify this.

The chiral case seems to be much simpler in this respect. There, we found
that the Jackiw-Rajaraman parameter $a_{c}$ does not play any distinctive
role in characterizing different regimes (what is consistent with several
other results in literature, that say, for example, that non-trivial
topology sectors do not contribute to the correlation functions in the CSM 
\cite{tiao-linhares}). The role of chiral gauge symmetry in this phenomenon
deserves more investigation.

Could we arrive at similar conclusions without performing the exact
renormalization? In the chiral model, the formal elimination of the UV
divergence allows one to get the correct information about the short (UV)
and long (IR) distances behavior. So, one would be tempted to say that, to
obtain this kind of information, renormalization is superfluous. This is not
true. Let us briefly comment on a previous paper on the ASM where
renormalization is not performed explicitly \cite{mitra}. The fermionic
propagator does not show the $Z_{\psi }$ wavefunction renormalization
constant (although the authors mention it). The formal elimination of the UV
divergence does not allow one to obtain information on the short (UV) and
long (IR) distances behavior because the massless boson propagator $D_{F}(x)$
has still a dependence in the infrared cut-off. So, it is premature to speak
of the model as ``non-confining''. Preliminary analysis, carried by us,
suggests two quite different phases for the anomalous Schwinger model,
according to the condition $a<1$ or $a>1$. If this analysis can be carried
somehow to four dimensions, it can give quite different perspectives to the
problem of confinement. This is being actively investigated by us and will
be reported elsewhere.

We are finally in position of computing ``physical'' (two dimensional)
amplitudes, like Compton scattering, for example. This would be very
important, to finally decide about the correct degree of arbitrariness
represented by the Jackiw-Rajaraman parameter $a$. To see the complete
effect of $a$ in the physical results we should be able to produce the
analog of a K\"{a}llen-Lehmann representation for the propagator of a theory
with dynamical mass generation, in order to produce a LSZ formula.

We could say that finally the denomination ``exactly soluble model'' makes
full sense. Only after the exact renormalization one can know, in principle,
all correlation functions of the model. For a more physical point of view, a
Hilbert space analysis of the ASM should be conducted in detail (the
corresponding analysis already exists for the CSM \cite{boyanovsky}). This
analysis would identify physical operators and the different sectors present
in the Hilbert space of the model, stating clearly what symmetries are
present and what are violated at quantum level. This and the other questions
mentioned in earlier paragraphs are actually under active investigation, and
results will be reported elsewhere.

{\large Acknowledgment: }{\it This work is part of the doctorate thesis of
R. Casana at CBPF. He is now supported by a CNPq post-doctoral fellowship at
IFT. We would like to dedicate this paper to the memory of our teacher and
friend Prof. Juan Alberto Mignaco.}

\end{document}